\documentclass[12pt]{article}
\usepackage{amsfonts}
\begin{document}
\title{\bf Higher Spin Particles with Bosonic Counterpart of Supersymmetry}
\author{{\bf Sergey Fedoruk}\thanks{On leave 
of absence from Ukrainian Engineering--Pedagogical Academy, Kharkov, Ukraine.}, \ {\bf Jerzy Lukierski}\thanks{Supported by KBN grant 1P03B01828.}
 \\ \\
\ ${}^{\ast}${\it\normalsize Bogoliubov Laboratory of
Theoretical Physics, JINR,} \\ 
{\it\normalsize 141980 Dubna, Moscow Region, Russian Federation}\\
{\normalsize {\it e-mail:} fedoruk@theor.jinr.ru}\\ 
\ ${}^{\dag}${\it\normalsize Institute for Theoretical Physics, University of Wroc{\l}aw} \\
{\it\normalsize pl. Maxa Borna 9, 50-204 Wroc{\l}aw, Poland}\\
{\normalsize {\it e-mail:} lukier@ift.wroc.pl}}
\date{}
\maketitle
\begin{abstract} 
We propose the relativistic point particle models invariant under the bosonic counterpart of SUSY. The particles move along the world lines in four dimensional Minkowski space extended by $N$ commuting Weyl spinors. The models provide after first quantization the non--Grassmann counterpart of chiral superfields, satisfying Klein--Gordon equation. Free higher spin fields obtained by expansions of such chiral superfields satisfy the $N=2$ Bargman--Wigner equations in massive case and Fierz--Pauli equations in massless case.
\end{abstract}

\noindent{\large\bf 1. Introduction.} 
Higher spin fields (see e.g.~\cite{Vas}-\cite{BekVas}) were investigated recently mainly due to their relations to string theory. For the description of higher spin fields the usual space--time is often extended by additional coordinates, e.g. commuting tensorial coordinates and/or commuting spinorial variables~\cite{Vas}-\cite{BBAST} having twistorial origin~\cite{PenMac}. Higher spin fields do appear as component fields in expansions of fields with respect to additional coordinate variables. It appears that the system of all higher spin fields possesses symmetry which is an extension of standard Poincare or conformal symmetries. In four dimensional space--time the system of massless higher spin fields has $Sp(8)$ symmetry or its supersymmetric extensions $OSp(N|8)$ ($N=1,2$) (see e.g.~\cite{IvLuk}).

In this report which is based on our paper~\cite{FedLuk} we propose new particle models invariant under bosonic counterpart of SUSY. The quantization of these particles produce infinite number of higher spin fields with all spins (helicities in massless case). The particle model with a trace of `bosonic' SUSY has been considered in~\cite{FedZim} for description of the relativistic particle with fixed spin (helicity). The realizations of `even' superalgebra was used also in~\cite{Lecht} for the description of spectrum of the critical open $N=2$ string in $2+2$ dimensions. 

The plan of our report is the following. In Sect. 2 we define the model describing the particle trajectory in the Minkowski space extended by $N$ Weyl commuting spinors. We determine the complete set of constraints and classify them. In Sect. 3 and 4 using Gupta--Bleuler method we perform the quantization of the models. The wave function describing first--quantized theory satisfies the Klein--Gordon equation and the bosonic counterpart of chirality condition. In expansion of wave function with respect to commuting spinors the component fields describe (anti)self-dual field strenghts and in massless case satisfy the Fierz-Pauli equations. It appears that in the case of bosonic counterpart of $N=2$ SUSY one can obtain also the linear Bargmann--Wigner equations for $D=4$ higher spin fields. In last section we shall summarize obtained results and present some unsolved questions related with our framework.
\\
\\
{\large\bf 2. Action with bosonic SUSY and the constraints.}
We describe the classical mechanics of higher spin particles by the following action
\begin{equation}\label{act}
S=\int d\tau\,{\cal L} \,, \qquad
{\cal L}=-{\textstyle\frac{1}{2e}}(\dot\omega_\mu\dot\omega^\mu+
e^2 m^2) -i
m(a_{ij}\dot\lambda^\alpha_i\lambda_{\alpha j} - \bar
a_{ij}\bar\lambda_{\dot\alpha
j}\dot{\bar\lambda}{}^{\dot\alpha}_i)\,.
\end{equation}
The action~(\ref{act}) describes propagation of
the particle in Minkowski space extended by commuting complex Weyl spinors coordinates $\lambda^\alpha_i(\tau)$,
$\bar\lambda_i^{\dot\alpha}=(\overline{\lambda^\alpha_i})$. We shall consider $N=2$ case ($i=1,2$) and $N=1$ case (no internal subindices). The constant matrix $a_{ij}$ is symmetric, $a_{ij}=a_{ji}$; if $a_{ij}=-a_{ji}$ the last terms in~(\ref{act}) are total derivatives because $a_{ij}\dot\lambda^\alpha_i\lambda_{\alpha j}=\frac{1}{2}
(a_{ij}\lambda^\alpha_i\lambda_{\alpha j})^{\cdot}$. The variable $e$ in Lagrangian~(\ref{act}) describes the einbein. Constant $m$ is the mass of the particle.

The $\omega$--form can be written in general case as follows
\begin{equation}\label{om}
\dot\omega^\mu=\dot x^\mu- i\kappa_{ij}
(\dot\lambda^\alpha_i\sigma^\mu_{\alpha\dot\beta}
\bar\lambda^{\dot\beta}_j-
\lambda^\alpha_j\sigma^\mu_{\alpha\dot\beta}
\dot{\bar\lambda}^{\dot\beta}_i)
\end{equation}
where constant matrix $\kappa_{ij}=\kappa_{ji}$ can be choose in the form
$\kappa_{ij}=\left(
\begin{array}{cc}
  1 & 0 \\
  0 & \kappa \\
\end{array}
\right)$ with real $\kappa$ by linear redefinitions of spinors
$\lambda^\alpha_i$ in $N=2$ internal space.

The action~(\ref{act}) is invariant
under the following spinorial bosonic transformation
\begin{equation}\label{trans}
\delta x^\mu=
i\kappa_{ij}(\lambda^\alpha_i\sigma^\mu_{\alpha\dot\beta}
\bar\varepsilon^{\dot\beta}_j-
\varepsilon^\alpha_i\sigma^\mu_{\alpha\dot\beta}
\bar\lambda^{\dot\beta}_j)\,,
\qquad \delta\lambda^\alpha_i=\varepsilon^\alpha_i \,, \qquad
\delta\bar\lambda^{\dot\alpha}_i=\bar\varepsilon^{\dot\alpha}_i
\end{equation}
where $\varepsilon^\alpha_i$ is a constant commuting Weyl spinors. Conserved Noether spinorial charges corresponding to the transformations~(\ref{trans}) are
\begin{equation}\label{R}
R_{\alpha i} \equiv \pi_{\alpha i}
-i\kappa_{ij}p_{\alpha\dot\beta}\bar\lambda^{\dot\beta}_j - i m a_{ij} \lambda_{\alpha j}\,, 
\qquad
\bar R_{\dot\alpha i}\equiv \bar\pi_{\dot\alpha i} + i\kappa_{ij}\lambda^\beta_j p_{\beta\dot\alpha} +im\bar a_{ij}\bar\lambda_{\dot\alpha j}
\end{equation}
where $p_\mu$, $\pi_{\alpha i}$, $\bar\pi_{\dot\alpha i}$ are the canonical momenta. Using the canonical Poisson brackets
\begin{equation}
\{x^\mu, p_\nu\}=\delta^\mu_\nu\,, \qquad\{\lambda^\alpha_i,
\pi_{\beta j}\}=\delta^\alpha_\beta\delta_{ij}\,, \qquad\{{\bar\lambda}^{\dot\alpha}_i,\bar\pi_{\dot\beta j}\}=
\delta^{\dot\alpha}_{\dot\beta}\delta_{ij}
\end{equation}
we obtain the PB algebra
\begin{equation}\label{al1-ch}
\{R_{\alpha i}, \bar R_{\dot\beta
j}\}=-2i\kappa_{ij}p_{\alpha\dot\beta}\,,
\quad\,\,\,\,
\{R_{\alpha i}, R_{\beta j}\}=2i
ma_{ij}\epsilon_{\alpha\beta}\,,\quad\,\,\,\, 
\{\bar R_{\dot\alpha i}, \bar
R_{\dot\beta j}\}= -2im\bar a_{ij} \epsilon_{\dot\alpha\dot\beta}
\end{equation}
which is classical (Poisson bracket) realization of bosonic counterpart of $N=2$ supersymmetry algebra with central charges $Z_{ij}=ma_{ij}$, $\bar Z_{ij}=m\bar a_{ij}$. Since the spinor variables are commuting and the Poisson brackets in~(\ref{al1-ch}) are even, the quantum realization of the algebra~(\ref{al1-ch}) is constructed in terms of the commutators
\begin{equation}\label{al1-q}
[ R_{\alpha i}, \bar R_{\dot\beta
j} ]=2\kappa_{ij}p_{\alpha\dot\beta}\,,
\qquad
[ R_{\alpha i}, R_{\beta j} ]=-2
ma_{ij}\epsilon_{\alpha\beta}\,,\qquad 
[ \bar R_{\dot\alpha i}, \bar
R_{\dot\beta j} ]= 2m\bar a_{ij} \epsilon_{\dot\alpha\dot\beta}
\end{equation}
in contrast to the algebra of anticommutators in standard $N=2$ supersymmetry.

The model~(\ref{act}) has the following nontrivial constraints (we omit the con\-stra\-int which im\-p\-lies pure gauge character of the einbein $e$)
\begin{equation}\label{T2}
T \equiv p^2 -m^2\approx 0\,,
\end{equation}
\begin{equation}\label{D2}
D_\alpha \equiv \pi_{\alpha i}
+i\kappa_{ij}p_{\alpha\dot\beta}\bar\lambda^{\dot\beta}_j+
ima_{ij}\lambda_{\alpha
j}\approx 0\,,
\qquad
\bar D_{\dot\alpha}\equiv \bar\pi_{\dot\alpha i} -
i\kappa_{ij}\lambda^\beta_j p_{\beta\dot\alpha} -im\bar a_{ij}
\bar\lambda_{\dot\alpha j}\approx 0\,.
\end{equation}

Nonvanishing Poisson brackets of the constraints~(\ref{T2})--(\ref{D2}) are 
\begin{equation}\label{al1-2}
\{D_{\alpha i}, \bar D_{\dot\beta
j}\}=2i\kappa_{ij}p_{\alpha\dot\beta}\,,
\quad\,\,\,\,
\{D_{\alpha i}, D_{\beta j}\}=-2i
ma_{ij}\epsilon_{\alpha\beta}\,,\quad\,\,\,\, \{\bar D_{\dot\alpha i}, \bar
D_{\dot\beta j}\}= 2im\bar a_{ij} \epsilon_{\dot\alpha\dot\beta}\,.
\end{equation}

The constraint~(\ref{T2}) $T\approx 0$ is a first class constraint. For classifying of the spinor constraints~(\ref{D2}) we look for the determinant of the matrix
\begin{equation}\label{C2}
{\cal C}=\left(
\begin{array}{cc}
  \{D_{\alpha i}, D_{\beta j}\} & \{D_{\alpha i}, \bar D_{\dot\beta j}\} \\
  \{\bar D_{\dot\alpha i}, D_{\beta j}\} & \{\bar D_{\dot\alpha i}, \bar D_{\dot\beta j}\} \\
\end{array}
\right)= \left(
\begin{array}{cc}
  -2i ma_{ij}\epsilon_{\alpha\beta} & 2i\kappa_{ij}p_{\alpha\dot\beta} \\
  -2i\kappa_{ij}p_{\beta\dot\alpha} & 2im\bar a_{ij} \epsilon_{\dot\alpha\dot\beta} \\
\end{array}
\right)\,.
\end{equation}

If matrix $(a_{ij})$ is diagonal it follows for $N=1,2$ that in massive case $\det{\cal C}$ is always nonzero, therefore all the constraints~(\ref{D2}) are of second class.

In case of antidiagonal matrix $(a_{ij})$ the matrix~(\ref{C2}) has vanishing determinant when $\kappa=-|a_{12}|^2<0$. Only in such a case the first class constraints are present in the model~(\ref{act}).\footnote{We note that in case of usual $N=2$ massive superparticle~\cite{AzLuk} when spinor variables
are Grassmannian and the matrix $(a_{ij})$ is skew--symmetric, the first class constraints are present if
$\kappa=|a_{12}|^2>0$.} Thus in massive case if we wish to have spinorial first class constraints we should consider $N\geq 2$ bosonic supersymmetry.

If $N=2$ we shall consider a simple choice $\kappa=-a_{12}=-1$.
In such a case the formulation~(\ref{act}) has an attractive interpretation if we pass to the commuting four--component Dirac spinor 
$
\psi_a= \left(
         \begin{array}{c}
           \lambda_{\alpha 1}\\
           {\bar\lambda}^{\dot\alpha}_2 \\
         \end{array}
       \right)
$, 
$\bar\psi^a=(\psi^+\gamma_0)^a=(\lambda^\alpha_2,
\bar\lambda_{\dot\alpha 1})$,
where $a=1,2,3,4$ is Dirac index. The Lagrangian~(\ref{act}) takes the simple form
\begin{equation}\label{Lagr2D}
{\cal L}=-{\textstyle\frac{1}{2e}}(\dot\omega_\mu\dot\omega^\mu+
e^2 m^2) -i
m(\dot{\bar\psi}\psi-{\bar\psi}\dot\psi)\,,
\end{equation}
\begin{equation}\label{om2D}
\dot\omega^\mu=\dot x^\mu+ i
(\dot{\bar\psi}\gamma^\mu\psi-{\bar\psi}\gamma^\mu\dot\psi)\,.
\end{equation}

In massless case ($m=0$) the matrix~(\ref{C2}) has vanishing determinant and even if $N=1$ the half of the spinorial constraints are first class.
\\
\\
{\large\bf 3. Gupta--Bleuler quantization of the model with $N=1$ bosonic SUSY.}
We shall perform the quantization
using Gupta--Bleuler technique what implies 
the split of the second class constraints into
complex--conjugated pairs, with holomorphic and antiholomorphic
parts forming separately the subalgebras of first class constraints. 

In \underline{massive $N=1$ case} the 
algebra~(\ref{al1-2}) of the constraints~(\ref{D2}) does not satisfy the Gupta--Bleuler requirements. However, the redefined constraints 
\begin{equation}\label{calD}
{\cal D}_\alpha=D_\alpha+
{\textstyle\frac{b}{m}}p_{\alpha\dot\beta}\bar D^{\dot\beta}\,,
\qquad \bar{\cal D}_{\dot\alpha}=\bar D_{\dot\alpha}+
{\textstyle\frac{b}{m}}D^{\beta}p_{\beta\dot\alpha}
\end{equation}
have the following algebra (we take $a_{11}=1$ without the loss of generality and we obtain $b=1\pm\sqrt{2}$)
\begin{equation}
\{{\cal D}_\alpha, {\cal D}_\beta \}=
{\textstyle\frac{2i}{m}}\epsilon_{\alpha\beta}T\,,\quad\,\,\, \{\bar{\cal
D}_{\dot\alpha}, \bar{\cal
D}_{\dot\beta}\}=-{\textstyle\frac{2i}{m}}
\epsilon_{\dot\alpha\dot\beta}T \,,\quad\,\,\, \{{\cal D}_\alpha, \bar
{\cal D}_{\dot\beta}\}=-8bip_{\alpha\dot\beta}
-{\textstyle\frac{2b^2i}{m^2}}p_{\alpha\dot\beta}T
\end{equation}
i. e. are suitable for application
of Gupta--Bleuler quantization method. The wave function which satisfies the Klein--Gordon constraint~(\ref{T2}) and spinorial wave equations ($\bar{\cal D}_{\dot\alpha}\Psi=0$ (chiral case) or ${\cal D}_{\alpha}\Psi=0$ (antichiral case)) provide the bosonic (non--Grassmann) counterpart of $D=4$ $N=1$ chiral superfield. It is possible to introduce new spinorial variables $\lambda^{\prime\alpha}$, $\bar\lambda^{\prime\dot\alpha}$, $\pi^\prime_\alpha$, $\bar\pi^\prime_{\dot\alpha}$ via canonical
transformation (see details in~\cite{FedLuk})
in which new constraints~(\ref{calD}) have the form
\begin{equation}\label{calD1}
{\cal D}_\alpha = \pi^\prime_\alpha
-4bip_{\alpha\dot\beta}\bar\lambda^{\prime\dot\beta}\approx
0\,,\qquad \bar{\cal D}_{\dot\alpha}= \bar\pi^\prime_{\dot\alpha}
+4bi\lambda^{\prime\beta} p_{\beta\dot\alpha}\approx 0\,.
\end{equation}
Solving chirality condition we obtain that the expansion of the wave function with respect to new spinorial variables contains infinite number space--time fields $\psi_{\alpha_1\cdots\alpha_n}(x)=
\psi_{(\alpha_1\cdots\alpha_n)}(x)$. They satisfy the Klein--Gordon equation ($\Box\equiv\partial_\mu\partial^\mu$)
\begin{equation}\label{sol1}
(\Box +m^2)\psi_{\alpha_1\cdots\alpha_n}(x)=0\qquad (n=1,2,\dots)\,.
\end{equation}
In antichiral case do appear in the expansion of the wave function the infinite number of fields $\bar\psi_{(\dot\alpha_1\cdots\dot\alpha_n)}(x)$ with dotted Weyl indices.

In \underline{massless $N=1$ case} the spinorial constraints 
\begin{equation}\label{Dm=0}
D_\alpha = \pi_\alpha +ip_{\alpha\dot\beta}\bar\lambda^{\dot\beta}
\approx 0\,,\qquad \bar D_{\dot\alpha}= \bar\pi_{\dot\alpha} -
i\lambda^\beta p_{\beta\dot\alpha} \approx 0
\end{equation}
are the mixture of first and second class constraints. The spinorial bosonic first class constraints are obtained from~(\ref{Dm=0}) by the multiplication with
$p_{\alpha\dot\beta}$: 
\begin{equation}
F^{\dot\alpha}=
p^{\dot\alpha\beta}D_\beta \approx 0\,, \qquad \bar
F^{\alpha}= \bar D_{\dot\beta}p^{\dot\beta\alpha} \approx 0\,.
\end{equation}
Unfortunately these constraints are reducible since 
\begin{equation}
p_{\alpha\dot\beta}F^{\dot\beta}\approx 0\, \qquad\bar F^{\beta}p_{\beta\dot\alpha}\approx 0\,.
\end{equation}
Irreducible separation of first and second class constraints is obtained by the projection of spinorial constraints~(\ref{Dm=0}) on spinors $\lambda^\alpha$ and $\bar\lambda_{\dot\alpha}p^{\dot\alpha\alpha}$. The constraints
\begin{equation}\label{cons-G}
G\equiv\lambda^{\alpha} D_\alpha \approx 0\,,\qquad\qquad \bar G
\equiv\bar D_{\dot\alpha}\bar\lambda^{\dot\alpha} \approx 0
\end{equation}
are second class whereas the constraints
\begin{equation}\label{cons-F}
F\equiv\bar\lambda_{\dot\alpha}p^{\dot\alpha\alpha} D_\alpha
\approx 0\,,\qquad\qquad \bar F \equiv\bar
D_{\dot\alpha}p^{\dot\alpha\alpha}\lambda_{\alpha} \approx 0
\end{equation}
are first class. 

Because the spinors $\lambda^\alpha$ and $\bar\lambda_{\dot\alpha} p^{\dot\alpha\alpha}$ are independent the pair of constraints $G\approx 0$ and $F\approx 0$ is equivalent to the constraints $D_{\alpha}\approx 0$. Similarly the constraints $\bar G\approx
0$ and $\bar F\approx 0$ are equivalent to the constraints $\bar D_{\dot\alpha}\approx 0$. Thus we have two sets of the wave equations: `bosonic chiral' case  
\begin{equation}
T\,\Psi =0\,,\qquad F\,\Psi =0\,,\qquad \bar D_{\dot\alpha}\,\Psi =0
\end{equation}
or `bosonic antichiral' one 
\begin{equation}
T\, \Psi =0\,,\qquad \bar F\,\Psi =0\,,\qquad D_{\alpha}\,\Psi =0\,.
\end{equation}

In the representation 
\begin{equation}
p_\mu=-i{\partial}_\mu\,, \qquad\pi_{\alpha}= -i{\partial}_\alpha\,, \qquad\bar\pi_{\dot\alpha}= -i\bar{\partial}_{\dot\alpha} 
\end{equation}
the equations in chiral case
$$
\Box\,\Psi=0\,, \qquad 
\bar D_{\dot\alpha}\,\Psi= (-i\bar\partial_{\dot\alpha} -
\lambda^\beta \partial_{\beta\dot\alpha})\,\Psi=0 \,, \qquad
-i\bar\lambda_{\dot\alpha}\partial^{\dot\alpha\alpha}
D_\alpha\,\Psi=
-\bar\lambda_{\dot\alpha}\partial^{\dot\alpha\alpha}
\partial_\alpha\,\Psi=0
$$
give only the dependence of the wave function on left--chiral variables
\begin{equation}
z_L^{}\equiv(x^\mu_L= x^\mu+i\lambda\sigma^\mu\bar\lambda,\,\lambda^\alpha)\,.
\end{equation}
We obtain the expansion
\begin{equation}\label{wf-exp}
\Psi(x^{}_L,\lambda)=\sum_{n=0}^{\infty}
\lambda^{\alpha_1}\ldots\lambda^{\alpha_n}
\phi_{\alpha_1\ldots\alpha_n}(x^{}_L)\,.
\end{equation}
The component fields are completely symmetric in spinor
indices,
$\phi_{\alpha_1\ldots\alpha_n}=\phi_{(\alpha_1\ldots\alpha_n)}$ and satisfy Fierz--Pauli equations for
component fields
\begin{equation}\label{PF}
\partial^{\dot\beta\beta}\,\phi_{\beta\alpha_2\ldots\alpha_n}=0\,.
\end{equation}
Scalar component field satisfies only the d'Alembert equation $\Box\,\phi =0$. The fields $\phi_{\alpha_1\ldots\alpha_n}(x)$ in the expansion of the wave function~(\ref{wf-exp}) are self--dual field strenghts of massless particles with helicities $n/2$. In antichiral case we obtain analogously anti-self--dual field strenghts of massless particles. 
\\
\\
{\large\bf 4. Quantum states describing particles with $N=2$ bosonic SUSY.}
The constraints~(\ref{T2})--(\ref{D2}) at $\kappa=-a_{12}=-1$, written
in Dirac notation, are the following
\begin{equation}\label{T-D}
T \equiv p^2 -m^2\approx 0\,, \qquad
D^a \equiv \pi^a +i\bar\psi^b(\hat p-m)_b{}^a\approx 0\,,
\qquad \bar D_a\equiv \bar\pi_a - i(\hat p-m)_a{}^b \psi_b \approx 0\,.
\end{equation}
Here $\pi^a$ and $\bar\pi_a$ are the conjugate momenta for $\psi_a$ and ${\bar\psi}^a$; its Poisson brackets are
\begin{equation}
\{\psi_a, \pi^b\}=\delta_a^b\, \qquad\{{\bar\psi}^a, \bar\pi_b\}=\delta^a_b
\end{equation}
where we use notation $\hat p\equiv \gamma^\mu p_\mu$.

From nonvanishing Poisson brackets of the constraints
\begin{equation}
\{\bar D_a, D^b \}=-2i(\hat p-m)_a{}^b
\end{equation}
we obtain directly that half of
the spinorial constraints~(\ref{T-D}) are first class
constraints. The projectors ${\cal P}_\pm\equiv \frac{1}{2m}(m\pm\hat p)$ define respectively the first class constraints
\begin{equation}\label{F}
F^a =D^b(\hat p+m)_b{}^a\,, \qquad \bar F_a =(\hat p+m)_a{}^b \bar
D_b
\end{equation}
and the second class
\begin{equation}\label{G}
G^a =D^b(\hat p-m)_b{}^a\,, \qquad \bar G_a =(\hat p-m)_a{}^b \bar
D_b\,.
\end{equation}
But due to the reducibility conditions
\begin{equation}
F^b(\hat p-m)_b{}^a\approx 0\, \qquad (\hat p-m)_a{}^b \bar F_b\approx 0
\end{equation}
if $T=p^2-m^2\approx 0$ the eight constraints
$(F^a, \bar F_a)$ has only four (real) independent constraints.
Analogously, the constraints $(G^a, \bar G_a)$ contain also four (real) independent constraints. 

In a way depending on the choice of second class constraints imposed on the wave function, we obtain (see details in~\cite{FedLuk}) that the wave function satisfies or the `bosonic chiral' equations 
\begin{equation}
T\,\Psi=0\,, \qquad \bar D_a\,\Psi =0\,, \qquad F^a\,\Psi =0
\end{equation}
or the `bosonic antichiral' ones 
\begin{equation}
T\,\Psi =0\,, \qquad D^a\,\Psi =0\,, \qquad\bar F_a\,\Psi =0\,.
\end{equation}

In `bosonic chiral' case the wave equations
\begin{equation}\label{eq-T,D,F}
(\Box^2 +m^2) \Psi=0\,,\qquad -i[\frac{\partial}{\partial\bar\psi^a}-(i\hat \partial+m)_a{}^b \psi_b ] \Psi=0\,, \qquad i\frac{\partial}{\partial\psi_b}(i\hat \partial -m)_b{}^a
\Psi=0
\end{equation}
have the general solution
\begin{equation}\label{Psi}
\Psi(x,\psi,\bar\psi) =e^{\bar \psi(i\hat\partial+m)\psi}\sum_{n=0}^\infty
\psi_{a_1}\cdots\psi_{a_n}\phi^{a_1\cdots a_n}(x)
\end{equation}
where the component fields $\phi^{a_1\cdots a_n}(x)$ are completely symmetric with respect to all Dirac indices,
$
\phi^{a_1\cdots a_n}(x)=\phi^{(a_1\cdots a_n)}(x)
$,
and satisfy the Dirac equations
\begin{equation}\label{BW}
(i\hat \partial -m)_{a_1}{}^{b}\phi^{a_1 a_2\cdots a_n}(x)=0\,.
\end{equation}
From~(\ref{BW}) follows the Klein--Gordon equation~(\ref{eq-T,D,F}).
Finally we have obtained the Bargmann--Wigner fields
describing massive particles of spins $n/2$.
\\
\\
{\large\bf 5. Conclusions.}
We have considered the models of the relativistic point particles propagating on fourdimensional Minkowski space extended by commuting Weyl spinors. The models are invariant under bosonic (non--Grassmann) counterpart of SUSY. The main results are the following:
\begin{itemize}
\item Higher spin fields emerge as the result of first quantization of the proposed models.
\item In massless case one obtained infinite set of field strenghts with all helicities satisfying linear Fierz--Pauli equations.
\item If we quantize the massive particle with $N=1$ bosonic counterpart of SUSY we obtain the massive free fields with any spin which satisfy only Klein--Gordon equation. We stress however that the first order equations of motions are missing.
\item For massive particle with $N=2$ bosonic counterpart of SUSY we get after quantization the wave function described by Bargmann--Wigner equations.
\end{itemize}

Let us note that some questions still should be answered. For instance, we do not understand the relation of our formalism with the unfolded formulation of higher spin fields by Vasiliev (see e.g.~\cite{Vas}) and link with the formulation using tensorial coordinates (see e.g.~\cite{BPST}). Also in our approach appears nonstandard relation between spin and statistics: both integer and half--integer spin fields have the same bosonic statistic. Here one should add that the analogous situation with statistics appears also in higher spin fields theory formulated on twistor spaces~\cite{PenMac}, \cite{FedZim-2tw}, \cite{BetAzLuk}.
\\
\\
{\large\bf Acknowledgments.} The authors would like to thank the Organizers of the International Workshop SQS'03 for their warm hospitality and additionally E.A.~Ivanov for fruitful discussions and valuable comments. One of the author (J.L.) would like to thank the EPSRC and University of Durham, where this report was completed.

\end{document}